\documentclass[11pt]{article}

\usepackage{graphicx}
\usepackage{amsmath}
\usepackage{amsfonts}
\usepackage{amssymb}
\usepackage{graphicx}

\renewcommand{\arraystretch}{1.3}

\catcode`\@=11
\def\marginnote#1{}

\newcount\hour
\newcount\minute
\newtoks\amorpm
\hour=\time\divide\hour by60
\minute=\time{\multiply\hour by60 \global\advance\minute by-\hour}
\edef\standardtime{{\ifnum\hour<12 \global\amorpm={am}%
        \else\global\amorpm={pm}\advance\hour by-12 \fi
        \ifnum\hour=0 \hour=12 \fi
        \number\hour:\ifnum\minute<10 0\fi\number\minute\the\amorpm}}
\edef\militarytime{\number\hour:\ifnum\minute<10 0\fi\number\minute}

\def\draftlabel#1{{\@bsphack\if@filesw {\let\thepage\relax
      \xdef\@gtempa{\write\@auxout{\string
          \newlabel{#1}{{\@currentlabel}{\thepage}}}}}\@gtempa \if@nobreak
    \ifvmode\nobreak\fi\fi\fi\@esphack} \gdef\@eqnlabel{#1}}
    \def\@eqnlabel{}
\def\@vacuum{}
\def\draftmarginnote#1{\marginpar{\raggedright\scriptsize\tt#1}}

\def\draft{
  \oddsidemargin -.5truein
  \def\@oddfoot{\footnotesize \sl preliminary draft \hfil
    \rm\thepage\hfil\sl\today\quad\militarytime}
  \let\@evenfoot\@oddfoot \overfullrule 3pt
    \let\label=\draftlabel
    \let\marginnote=\draftmarginnote
  \def\@eqnnum{(\theequation)\rlap{\kern\marginparsep\tt\@eqnlabel}%
    \global\let\@eqnlabel\@vacuum}

  }

\makeatletter
\newdimen\normalarrayskip              
\newdimen\minarrayskip                 
\normalarrayskip\baselineskip
\minarrayskip\jot
\newif\ifold             \oldtrue            \def\new{\oldfalse}
\def\arraymode{\ifold\relax\else\displaystyle\fi} 
\def\eqnumphantom{\phantom{(\theequation)}}     
\def\@arrayskip{\ifold\baselineskip\z@\lineskip\z@
     \else
     \baselineskip\minarrayskip\lineskip2\minarrayskip\fi}
\def\@arrayclassz{\ifcase \@lastchclass \@acolampacol \or
\@ampacol \or \or \or \@addamp \or
   \@acolampacol \or \@firstampfalse \@acol \fi
\edef\@preamble{\@preamble
  \ifcase \@chnum
     \hfil$\relax\arraymode\@sharp$\hfil
     \or $\relax\arraymode\@sharp$\hfil
     \or \hfil$\relax\arraymode\@sharp$\fi}}
\def\@array[#1]#2{\setbox\@arstrutbox=\hbox{\vrule
     height\arraystretch \ht\strutbox
     depth\arraystretch \dp\strutbox
     width\z@}\@mkpream{#2}\edef\@preamble{\halign
\noexpand\@halignto
\bgroup \tabskip\z@ \@arstrut \@preamble \tabskip\z@ \cr}%
\let\@startpbox\@@startpbox \let\@endpbox\@@endpbox
  \if #1t\vtop \else \if#1b\vbox \else \vcenter \fi\fi
  \bgroup \let\par\relax
  \let\@sharp##\let\protect\relax
  \@arrayskip\@preamble}
\def\eqnarray{\stepcounter{equation}%
              \let\@currentlabel=\theequation
              \global\@eqnswtrue
              \global\@eqcnt\z@
              \tabskip\@centering
              \let\\=\@eqncr

 \halign to \displaywidth\bgroup
    \eqnumphantom\@eqnsel\hskip\@centering
    $\displaystyle \tabskip\z@ {##}$%
    \global\@eqcnt\@ne \hskip 2\arraycolsep
         $\displaystyle\arraymode{##}$\hfil
    \global\@eqcnt\tw@ \hskip 2\arraycolsep
         $\displaystyle\tabskip\z@{##}$\hfil
         \tabskip\@centering
    &{##}\tabskip\z@\cr}
\newfont{\hr}{msbm10}
\newfont{\ams}{msam10}

\hoffset= - 0.9in         

\def\beq{\begin{equation}}
\def\eeq{\end{equation}}
\def\ba{\beq\new\begin{array}{c}}
\def\ea{\end{array}\eeq}

\def\be{\ba}
\def\ee{\ea}

\def\N2{${\cal N}=2$}

\def\1N{${\cal N}=1$}
\def\4N{${\cal N}=4$}
\def\nn{\nonumber}

\def\p{\partial}

\newdimen\linethick  \linethick=0.4pt
\newdimen\hboxitspace    \hboxitspace=5pt
\newdimen\vboxitspace    \vboxitspace=5pt

\def\fr#1{%
\beq\new
\vcenter{
\hrule height\linethick
          \hbox{\vrule width\linethick
                \kern\hboxitspace
                \vbox{\kern\vboxitspace
                      \hbox{$\begin{array}{c}\displaystyle#1
         \end{array}$}%
                      \kern\vboxitspace}%
                \kern\hboxitspace
                \vrule width\linethick}%
          \hrule height\linethick}%
\eeq}

\renewcommand{\tt}[1][mer]{\hbox{\tiny{#1}}}

\newcommand{\Tr}{\mathop{\rm Tr}\nolimits}

\def\p{\partial}
\def\tr{{\rm tr}\,}
\def\Tr{{\rm Tr}\,}

\def\l[{\phantom.[}

\def\K{h}

\hoffset= - 1.0in         

\newdimen\linethick  \linethick=0.4pt
\newdimen\hboxitspace    \hboxitspace=5pt
\newdimen\vboxitspace    \vboxitspace=5pt

\def\fr#1{%
\beq\new
\vcenter{
\hrule height\linethick
          \hbox{\vrule width\linethick
                \kern\hboxitspace
                \vbox{\kern\vboxitspace
                      \hbox{$\begin{array}{c}\displaystyle#1
         \end{array}$}%
                      \kern\vboxitspace}%
                \kern\hboxitspace
                \vrule width\linethick}%
          \hrule height\linethick}%
\eeq}

\title{{\bf Character expansion
for HOMFLY polynomials. I. Integrability and difference equations } \vspace{.2cm}}
\author{{\bf A.Mironov}\footnote{ {\small {\it
Lebedev Physics Institute} and {\it ITEP, Moscow, Russia}};
mironov@itep.ru; mironov@lpi.ru}, {\bf A.Morozov}\thanks{{\small
{\it ITEP, Moscow, Russia}}; morozov@itep.ru}, {\bf
And.Morozov}\thanks{{\small {\it Moscow State University} and
{\it ITEP Moscow, Russia}};
Andrey.Morozov@itep.ru}\date{ }}

\begin{document}
 \maketitle

\vspace{-6.5cm}

\begin{center}
\hfill FIAN/TD-19/11\\
\hfill ITEP/TH-62/11\\
\end{center}

\vspace{5cm}

\centerline{ABSTRACT}

\bigskip

{\footnotesize We suggest to associate with each knot the set of
coefficients of its HOMFLY polynomial expansion into the Schur
functions. For each braid representation of the knot these
coefficients are defined unambiguously as certain combinations of
the Racah symbols for the algebra $SU_q$. Then, the HOMFLY
polynomials can be extended to the entire space of time-variables.
The so extended HOMFLY polynomials are no longer knot invariants,
they depend on the choice of the braid representation, but instead
one can naturally discuss their explicit integrable properties. The
generating functions  of
torus knot/link coefficients are turned to satisfy the Pl\"ucker
relations and can be associated with $\tau$-function of the KP
hierarchy, while generic knots correspond to more involved
systems. On the other hand, using the expansion into the Schur
functions, one can immediately derive difference equations
(A-polynomials) for knot polynomials which play a role of the string
equation. This adds to the previously demonstrated  use of these
character decompositions for the study of $\beta$-deformations from
HOMFLY to superpolynomials.}

\vspace{1.5cm}

\begin{flushright}
{\bf To the memory of Max Kreuzer}
\end{flushright}

\vspace{1.5cm}

\section{Introduction}

Knot theory is a very old and complicated mathematical domain, with
many deep ideas and results. Its counterpart in quantum
field theory is the $3d$ Chern-Simons (CS) model \cite{CS} and its
extensions to higher dimensions. For string/M-theory of
main interest are various Wilson averages in CS theory and, most
important, relations between them. These Wilson averages are known
in CS theory as HOMFLY "polynomials" \cite{HOMFLY}, while some of
the relations (some linear ones) have appeared under the name of
"quantum A-polynomial" \cite{Apols}. Knot theory is becoming
especially interesting today, because there is now a strong belief
that the HOMFLY polynomials are closely related to KP/Toda
$\tau$-functions, providing a minor deformation of these, while the
linear relations provide likewise minor deformations of the string
equations and the Virasoro constraints \cite{UFN3}. Since by now a
lot is known on the knot {\it phenomenology}, e.g. concrete HOMFLY
polynomials are easily available, say, at \cite{katlas,II}, the time
is coming to proceed to a {\it theoretical} analysis of the {\it
problem}, which can be, {\it in particular}, nicknamed as
$$
\boxed{ {\bf HOMFLY\ polynomials\ as\ deformed\ matrix\ model\
\tau-functions} }\ \footnote{We remind that the matrix model
$\tau$-functions is a particular class of $\tau$-functions which, in
addition to the bilinear Hirota equations, also satisfy a linear
"string equation" generating together with the Hirota equations an
entire set of Virasoro like constraints. These constraints can be
also described in terms of the AMM/EO topological recursion
\cite{AMMEO}. Sometimes (e.g. in $\beta$-ensembles) only these
Virasoro like constraints are known, while bilinear equations (and
associated Harer-Zagier recursion) remain to be found \cite{MMS}. It
seems that in knot theory we also typically know only linear
equations (A-polynomial). In this paper we find the bilinear identities, at least, in the
case of torus knots.}
$$
There are already numerous attempts in the literature, targeted at
this problem (perhaps, not formulated so explicitly), see, for
example, \cite{Mari}.

Our suggestion to attack the problem, after it is explicitly
formulated, is to rely upon the known common property of the
polynomial KP $\tau$-functions and HOMFLY polynomials: {\bf these
both can be expanded into the Schur functions}, the characters of
the linear group $GL(\infty)$. The Schur functions $S_Q\{p\}$ depend
on an infinite set of "time-variables" $p_k = kt_k$, $k=1,2,\ldots$
($p$ and $t$ are the two standard choices, widely used in different
fields), and correspond to the representations of the linear group
or, simply, are labeled by the Young diagrams
$Q=\{\lambda_1\geq\lambda_2\geq \ldots\geq 0\}$.

The KP $\tau$-functions (solutions to the bilinear Hirota equation)
are linear combinations of the Schur functions
(see \cite{UFN3,intcharexp} for reviews in the terms relevant for our purposes):
\be
\tau\{p\,|g\} =
\sum_Q g_Q S_Q\{p\,\}, \label{tauexpan}
\ee
provided the
coefficients $g_Q$ satisfy the infinite set of bilinear Pl\"ucker
relations,
\be
g_{[22]}g_{[0]} - g_{[21]}g_{[1]} + g_{[2]}g_{[11]} = 0,\nn
\ee

\begin{center}
\rule{7cm}{.4pt}
\end{center}
\be
g_{[32]}g_{[0]} - g_{[31]}g_{[1]} + g_{[3]}g_{[11]} = 0,  \nn\\
g_{[221]}g_{[0]} - g_{[211]}g_{[1]} + g_{[2]}g_{[111]} = 0,
\ee

\begin{center}
\rule{7cm}{.4pt}
\end{center}
\be\label{Plure}
g_{[42]}g_{[0]} - g_{[41]}g_{[1]} + g_{[4]}g_{[11]} = 0, \\
g_{[33]}g_{[0]} - g_{[31]}g_{[2]} + g_{[3]}g_{[21]} = 0, \\
g_{[321]}g_{[0]} - g_{[311]}g_{[1]} + g_{[3]}g_{[111]} = 0, \\
g_{[222]}g_{[0]} - g_{[211]}g_{[11]} + g_{[21]}g_{[111]} = 0, \\
g_{[2211]}g_{[0]} - g_{[2111]}g_{[1]} + g_{[2]}g_{[1111]} = 0,\\
\ldots
\ee
and the (infinite) set of coefficients $g =
\{g_Q\}$ describes a point in the Universal Grassmannian
\cite{uniGra}.
Moreover, one further generalizes $\tau$ in (\ref{tauexpan}) to be a
Toda-lattice $\tau$-function, provided the coefficients $g_R$
themselves depend on another infinite set of time variables, $\bar
p_k$, and
\be
g_Q = \sum_R g_Q^R S_R\{\bar p\} \label{vexpan}
\ee
where $g_Q^R$ satisfy some more involved bilinear relations. For
matrix model like $\tau$-functions, which {\it actually} arise in
the role of generating functions in quantum field theory, these
coefficients also satisfy {\it linear} relations, known as string
equations or, more generally, Virasoro constraints.

The HOMFLY polynomial
is equal to the properly defined\footnote{See \cite{MSm,3dAGT,DBSS} for recent review
of {\it existing} problems.}
Wilson loop average in CS theory with the group
$SU(N)$ and the coupling constant $q = \exp\left(\frac{2\pi
i}{k+N}\right)$:
\be
H_R^{\cal K} = \ \left<\tr_R P\exp\oint_{{\cal K}} {\cal
A}\right>_{CS(N,q)}
\ee
Usually the $N$-dependence is traded for
$A$-dependence, where $A=q^N$. Then $H_R$ is a polynomial in $A$
(modulo some common power of $A$ that depends on the normalization).
It is  labeled by the representation index $R$ and already in this
respect resembles the coefficients $g_R$ in (\ref{tauexpan}), only
the role of point of the Universal Grassmannian is now played by the
triple $({\cal K},A,q)$. However, for $q\neq 1$ these $H_R^{\cal K}$
do {\it not} satisfy the Pl\"ucker relations (\ref{Plure}),\footnote{
For instance, consider
the particular case of the HOMFLY polynomial at $N=2$, i.e. $A=q^2$. Then,
for the spin $j$ representation this knot polynomial is the Jones polynomial
$J_{2j+1}$ and the simplest Pl\"ucker relation in (\ref{Plure})
looks like
\be
g_{[0]}=J_1({\cal K})=1,\ \ \ \ \ g_{[1]}=J_2({\cal K}),\ \ \ \ \
g_{[2]}=J_3({\cal K}),\ \ \ \ \
g_{[11]}=J_1({\cal K})=1,\ \ \ \ \ \ g_{[21]}=J_2({\cal K}),\ \ \ \ \ \
g_{[22]}=J_1({\cal K})=1,\\
g_{[22]}g_{[0]} - g_{[21]}g_{[1]} + g_{[2]}g_{[11]} = 1-J_2^2({\cal K})+J_3({\cal K})\nn
\ee
At the same time, from the relation $J_{R^{\otimes m}}({\cal K})
=J_R({\cal K}^m)$, where ${\cal K}^m$ denotes $m$-cabling of the knot ${\cal K}$
(see, e.g., \cite[eq.(1.5b)]{DG}) it follows that
 $1+J_3({\cal K})=J_2({\cal K}^2)\ne
\Big(J_2({\cal K})\Big)^2$ (unless ${\cal K}$ is unknot).
Therefore, already the first relation in (\ref{Plure}) is not fulfilled.
} thus the
generating function
\be
\mathfrak{H}\{p|{\cal K}\} = \sum_R
H_R^{\cal K} S_R\{p\} \label{hexpan}
\ee
is some $q$-deformation of
the KP $\tau$-function, which still remains to be investigated and
understood.

However, the HOMFLY polynomials themselves possess another
expansion, similar to (\ref{vexpan}):
\be \boxed{ H_R^{\cal K}  = \sum_Q \K_R^Q S_Q\{p^*\} \equiv {\cal
H}_R\{p^*|{\cal K}\} } \label{HOexpan}
\ee
and relations (linear and non-linear) between the ${\cal
K}$-dependent coefficients $\K_R^Q$ are the ones to be found.
Like (\ref{tauexpan}) and like (a very different)
Vassiliev-Kontsevich expansion in knot theory, (\ref{HOexpan})
separates dependencies on different variables, in this case on the
group (which is contained in the time-variables) and on the knot,
which are strongly mixed in the original definitions (either through
CS theory or directly through braid representations, for an overview
of their still obscure connection see \cite{MSm}). An important
difference from (\ref{tauexpan}) and (\ref{vexpan}) is that $p^*$ in
(\ref{HOexpan}) is not an {\it arbitrary} point in the space of
time-variables: it is constrained to just a $2$-dimensional slice
\be
p^*_k = \frac{A^k-A^{-k}}{q-q^{-1}} = \frac{\{A^k\}}{\{q\}}
\label{p*slice}
\ee
Hereafter, we introduced a useful notation
$\{x\} = x - x^{-1}$ to simplify the formulas. For $A=q^N$ these
$p^*_k = [N]_q\equiv \{q^N\}/\{q\}$.

The manifest expressions for the Schur functions $S_Q\{p^*\}$ in
these special points (\ref{p*slice})
 are quite simple and generalize the standard hook formula \cite{hook}:
\be\label{he}
S_Q\{p^*\}= \prod_{(i,j)\in Q}
\frac{\{Aq^{i-j}\}}{\{q^{h_{i,j}}\}} \ \ \ \
\stackrel{A=q^N}{\longrightarrow}\ \ \ \ { \prod_{(i,j)\in Q}
\frac{[N+i-j]_q}{[h_{i,j}]_q} }
\ee
where $h_{i,j}$ is the hook
length.
\begin{figure}
\unitlength 1mm 
\linethickness{0.4pt}
\ifx\plotpoint\undefined\newsavebox{\plotpoint}\fi 
\begin{picture}(58.968,38.022)(-45,23)
\put(31.849,31.008){\line(0,1){27.014}}
\put(31.849,58.022){\line(1,0){3.048}}
\put(34.897,58.022){\line(0,-1){27.014}}
\put(34.897,31.008){\line(-1,0){3.048}}
\put(31.849,54.868){\line(1,0){3.0482}}
\put(31.954,51.715){\line(1,0){2.838}}
\put(34.897,31.113){\line(1,0){24.071}}
\put(58.968,31.113){\line(0,1){3.048}}
\put(58.968,34.161){\line(-1,0){27.119}}
\put(37.42,30.903){\line(0,-1){.2102}}
\put(37.63,31.113){\line(0,1){15.031}}
\put(37.63,46.144){\line(-1,0){5.6761}}
\put(31.954,36.999){\line(1,0){23.545}}
\put(55.499,36.999){\line(0,-1){5.781}}
\put(31.954,39.837){\line(1,0){18.079}}
\put(50.033,31.218){\line(0,1){8.514}}
\put(52.661,36.684){\line(0,-1){5.466}}
\put(32.059,42.886){\line(1,0){11.983}}
\put(44.042,42.886){\line(0,-1){11.878}}
\put(46.985,39.417){\line(0,-1){8.304}}
\put(40.678,42.57){\line(0,-1){11.352}}
\put(31.849,48.877){\line(1,0){2.9431}}
\put(42.15,28.906){\makebox(0,0)[cc]{$i$}}
\put(29.326,35.738){\makebox(0,0)[cc]{$j$}}
\put(42.15,35.528){\makebox(0,0)[cc]{{\bf x}}}
\qbezier(45.303,35.528)(55.972,41.677)(54.238,35.423)
\qbezier(42.465,38.261)(49.613,42.728)(42.465,41.519)
\put(54.763,39.417){\makebox(0,0)[cc]{$k$}}
\put(47.616,42.15){\makebox(0,0)[cc]{$l$}}
\end{picture}
\caption{The figure which illustrates the notation in the generalization of the standard
hook formula to the quantum dimensions (\ref{he}). Here the cross {\bf x} corresponds
to the box of the Young diagram with coordinates $(i,j)$. The corresponding hook
length is equal to $h_{i,j}=k+l+1$.}
\end{figure}
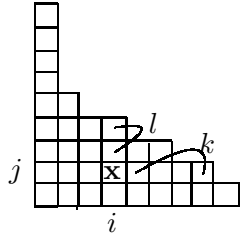

Given (\ref{HOexpan}), one can easily continue ${\cal H}_R\{
p|{\cal K}\}$ to arbitrary values of $p$, where it can be
compared with KP/Toda $\tau$-functions. The problem is, however, to
define the coefficients $\K_R^Q$. For most knots (represented by
braids with more than $3$ strands) they can not be obtained from the
known expressions for HOMFLY polynomials: for $|Q|\geq 4$ the $S_Q^*
= S_Q\{p^*\}$ form  a linearly dependent set of functions of the
$A$-variable (while $q$-dependence does not help, since $\K_Q$ can
also depend on $q$). In \cite{DBMMSS} we suggested to overcome this
problem by considering the HOMFLY polynomials for series of knots at
once and with the $\beta$-deformation \cite{betadef} switched on
(i.e. the "superpolynomials") \cite{superpolynomials}: then the
decomposition like (\ref{HOexpan}) becomes unambiguous and all the
coefficients $\K_R^Q$ can be found. This is a very promising and
interesting direction. However, there is an alternative approach
directly based on (quantum) group theory
underlying the CS theory and using the Reshetikhin-Turaev construction for the
HOMFLY polynomials \cite{RT} which arises in the temporal gauge, $A_0=0$
and which we basically exploit here.
The details of the approach can be found in a separate
paper \cite{II}, here we only briefly describe the scheme in section 2.

The remaining two sections describe two immediate applications of the character
expansion of the HOMFLY polynomials.
In section 3
we discuss integrable properties of various knots and explain that
in the case of torus knots the generating function of ${\cal H}_R\{\bar
p|{\cal K}\}$ is a KP $\tau$-function in variables $\bar t_k=p_k/k$:
\be
\tau\{t|{\cal K}\}=\sum_RS_R(p){\cal H}_R\{
p|{\cal K}\}
\ee
and similarly for the torus knots, while for other knots the
situation is not that simple. As usual, the concrete solution to the KP
equations is specified by the string equation (more generally, by
Virasoro/W like constraints). The role of this kind of equation for the
knot polynomials is played by the difference equations (A-polynomials) which we
discuss in the simplest case of the Jones polynomial in section 4.
We end in section 5 with contains some concluding remarks.

\section{HOMFLY polynomials as sums of characters}

\paragraph{Character expansion of HOMFLY polynomials.}
Here we outline only the basic idea, details and explanations are given in
\cite{II} and forthcoming papers of the series.
With a braid representation
of the knot we associate the character expansion of the colored
HOMFLY polynomial, i.e. represent
it as a linear combination of the Schur functions
(i.e. the $SU(\infty)/S(\infty)$ characters).
Such an expansion depends on the choice of a braid realization,
thus, its coefficients by themselves
are not knot invariants,
instead they are pure group theory quantities and
possess many nice properties.
For an $m$-strand braid ${\cal B}$ the HOMFLY polynomial in
representation $R$ is expanded as
\be\label{10}
H_R^{\cal B} = \tr_{R^{\otimes m}}\Big((q^\rho)^{\otimes m}{\cal B}\Big)
= \sum_{Q\vdash m|R|} h_R^Q[{\cal B}]\,S_Q^*(A)
\ee
where
\be
S_Q^*(A) = \tr_{R^{\otimes m}} (q^\rho)^{\otimes m}
= S_Q\{p_k^*\}, \ \ \ \ \ \ \ \ \ \
\ee
are the quantum dimensions of representations $Q$ of $SU(N)$.
Coefficients $h_R^Q[{\cal B}]$ do not depend on $A$,
i.e. on $N$,
thus, they can be evaluated from analysis of arbitrary group $SU_q(N)$.

Instead, these coefficients can be represented as traces in auxiliary
spaces of intertwiner operators ${\cal M}^Q_{R^m}$, whose dimension
is the number ${\rm dim} {\cal M}^Q_{R^m} = N_{R^m}^Q$ of times
the irreducible representation $Q$ appears in the $m$-th tensor power
of the representation $R$,
\be
R^{\otimes m} = \sum_{Q\vdash m|R|} {\cal M}_{R^m}^Q \otimes Q
\label{decoRmQ}
\ee
These new traces (which we denote $\Tr$ in order to differ from the traces
$\tr$ in the space of representation) are taken of products of diagonal quantum ${\cal R}$-matrices
$\widehat{\cal R}$ acting in ${\cal M}^Q_{R^m}$
and the "mixing matrices" intertwining the ${\cal R}$-matrices,
acting on different pairs of adjacent strands in the braid.
These mixing matrices, in their turn, can be represented as products
of universal constituents, associated with a switch between two
"adjacent" trees, describing various decompositions (\ref{decoRmQ}).

\paragraph{HOMFLY for any knot with 2,3,4 braids in the fundamental representation.}
In \cite{II} we exhaustively described
such representations for the coefficients $h_R^Q[{\cal B}]$
for {\it arbitrary} $m=2,3,4$-strand braids
and for the simplest representation $R=[1]$:
\be\label{13}
\underline{m=2}, \hspace{2cm} {\cal B} = {\cal R}^a: \\
H_{[1]}^{(a)} = q^a S_2^*(A) + \left(-\frac{1}{q}\right)^a S_{11}^*(A)
=  q^a S_2^*(A) \ +\ \left( q \longrightarrow -\frac{1}{q}\right)
\ee

\be
\underline{m=3}, \hspace{1cm} {\cal B} = ({\cal R}\otimes I)^{a_1}(I\otimes {\cal R})^{b_1}
({\cal R}\otimes I)^{a_2}(I\otimes {\cal R})^{b_2} \ldots :\\
H_{[1]}^{(a_1,b_1,a_2,b_2,\ldots)} =
q^{\sum_i (a_i+b_i)} S_3^*(A) + \left(-\frac{1}{q}\right)^{\sum_i(a_i+b_i)} S_{111}^*(A)
+ \Big(\Tr_{2\times 2} \widehat{\cal R}_2^{a_1}U_2 \widehat{\cal R}_2^{b_1}U_2^\dagger
 \widehat{\cal R}_2^{a_2}U_2 \widehat{\cal R}_2^{b_2}U_2^\dagger \ldots\Big) S_{21}^*(A)
\ee

Thus, an arbitrary $3$-strand braid is parameterized by a sequence of integers
$a_1,b_1,a_2,b_2,\ldots$, their meaning can be understood from the picture
(in this figure $a_1=-2$, $b_1=2$, $a_2=-1$, $b_2=3$:
this is knot $8_{10}$ ):

\vspace{0.7cm}

\unitlength 1mm 
\linethickness{0.4pt}
\ifx\plotpoint\undefined\newsavebox{\plotpoint}\fi 
\begin{picture}(145.5,53)(0,0)
\put(19.5,34.5){\line(1,0){13.25}}
\put(41.25,43.25){\line(1,0){11.25}}
\put(19.25,43){\line(1,0){13.25}}
\put(38.75,35){\line(1,0){13.75}}
\put(61.25,43.25){\line(1,1){8.75}}
\put(70,52){\line(1,0){14.75}}
\put(18.5,52){\line(1,0){41}}
\multiput(59.5,52)(.033505155,-.043814433){97}{\line(0,-1){.043814433}}
\put(58.25,35.25){\line(1,0){33.75}}
\multiput(92,35.25)(.033505155,.038659794){97}{\line(0,1){.038659794}}
\multiput(64.5,45)(.03289474,-.04605263){38}{\line(0,-1){.04605263}}
\put(65.75,43.25){\line(1,0){19}}
\multiput(84.5,43.5)(.0346153846,.0336538462){260}{\line(1,0){.0346153846}}
\multiput(84.75,52)(.03370787,-.03651685){89}{\line(0,-1){.03651685}}
\multiput(52.5,43)(.033653846,-.046474359){156}{\line(0,-1){.046474359}}
\multiput(52.5,35)(.03353659,.03353659){82}{\line(0,1){.03353659}}
\multiput(56.75,39)(.035447761,.03358209){134}{\line(1,0){.035447761}}
\multiput(32.25,43)(.033602151,-.041666667){186}{\line(0,-1){.041666667}}
\multiput(32.75,34.75)(.03333333,.03333333){75}{\line(0,1){.03333333}}
\put(37,39){\line(1,1){4.25}}
\put(99.75,35.25){\line(1,0){45.75}}
\multiput(100,35.5)(-.0336990596,.0352664577){319}{\line(0,1){.0352664577}}
\multiput(97.25,41)(.0336363636,.04){275}{\line(0,1){.04}}
\put(106.5,52){\line(1,0){7.75}}
\put(121.25,44){\line(1,0){6.75}}
\put(128,44){\line(5,6){7.5}}
\put(135.5,53){\line(1,0){8.25}}
\put(93.25,52.25){\line(1,0){5.75}}
\multiput(99,52.25)(.03353659,-.04268293){82}{\line(0,-1){.04268293}}
\multiput(103,47)(.03333333,-.05){60}{\line(0,-1){.05}}
\put(105,44){\line(0,1){0}}
\put(105,44){\line(1,0){9.5}}
\multiput(114.5,44)(.033632287,.036995516){223}{\line(0,1){.036995516}}
\put(122,52.25){\line(1,0){5.25}}
\multiput(127.25,52.25)(.03353659,-.03963415){82}{\line(0,-1){.03963415}}
\multiput(131.5,47)(.03333333,-.04166667){60}{\line(0,-1){.04166667}}
\put(133.5,44.5){\line(1,0){10.75}}
\multiput(114.25,52.25)(.03370787,-.03651685){89}{\line(0,-1){.03651685}}
\multiput(121,44)(-.03333333,.04666667){75}{\line(0,1){.04666667}}
\end{picture}

\vspace{-2.7cm}

Similarly,
\be\label{15}
\underline{m=4}, \hspace{1cm} {\cal B} =
({\cal R}\otimes I\otimes I)^{a_1}(I\otimes {\cal R}\otimes I)^{b_1}
 ({\cal R}\otimes I\otimes I)^{c_1}
({\cal R}\otimes I\otimes I)^{a_2}(I\otimes {\cal R}\otimes I)^{b_2}
({\cal R}\otimes I\otimes I)^{c_2} \ldots :
\\
H_{[1]}^{(a_1,b_1,c_1,a_2,b_2,c_2,\ldots)} =
q^{\sum_i (a_i+b_i+c_i)} S_4^*(A) + \left(-\frac{1}{q}\right)^{\sum_i(a_i+b_i+c_i)} S_{1111}^*(A) +
\\
+ \Big(\Tr_{2\times 2} \widehat{\cal R}_2^{a_1}U_2 \widehat{\cal R}_2^{b_1}U_2^\dagger
 \widehat{\cal R}_2^{c_1+a_2}U_2 \widehat{\cal R}_2^{b_2}U_2^\dagger  \widehat{\cal R}_2^{c_2+a_3}
 \ldots\Big) S_{22}^*(A) +
\\
+ \left\{\Big(\Tr_{3\times 3} \widehat{\cal R}_3^{a_1}U_3 \widehat{\cal R}_3^{b_1} V_3U_3\widehat{\cal R}_3^{c_1}
U_3^\dagger V_3^\dagger U_3^\dagger
\widehat{\cal R}_3^{a_2}U_3 \widehat{\cal R}_3^{b_2} V_3U_3\widehat{\cal R}_3^{c_2}
U_3^\dagger V_3^\dagger U_3^\dagger
 \ldots\Big) S_{31}^*(A)\ + \ \left(q \longrightarrow -\frac{1}{q}\right)\right\}
\ee
In these formulas:
\be
\hat{\cal R}_2 = \left(\begin{array}{cc} q & \\ &-\frac{1}{q}\end{array}\right)\ \ \ \ \ \ \ \ \ \
\hat{\cal R}_3 = \left(\begin{array}{ccc} q & & \\ & q &\\ &&-\frac{1}{q}\end{array}\right)
\ee
\be
U_2 = \left(\begin{array}{cc} c_2 & s_2\\ -s_2 &c_2\end{array}\right)\ \ \ \ \ \ \ \
U_3 = \left(\begin{array}{ccc} 1&& \\ &c_2 & s_2\\ &-s_2 &c_2\end{array}\right)\ \ \ \ \ \ \ \
V_3 = \left(\begin{array}{ccc} c_3 & s_3 & \\ -s_3 &c_3 & \\ && 1\end{array}\right)
\ee
Subscripts refer to the size of the matrices, the entries of rotation matrices $U$ and $V$
are given by
\be
c_k = \frac{1}{[k]}\,, \ \ \ \
s_k = \sqrt{\,1-c_k^2\,} = \frac{\sqrt{\,[k-1]^{\phantom {1^1}}\!\!\![k+1]\,}}{[k]}
\ee
These formulas provide a very transparent and convenient representation for infinitely
many HOMFLY polynomials
and seem to be very useful for any theoretical analysis of their general properties,
from integrability to linear Virasoro like relations
(including $A$-polynomials, spectral curves,
AMM/EO topological recursion etc). They describe in a very effective way the HOMFLY
polynomials' dependence on particular $a_i,b_i,c_i$, i.e. on the shape of the braid.
Therefore, further insights are important about the structure of these formulas
and their generalizations
(in \cite{II} the $m=5$ case is also investigated,
and the general formula for the coefficients $h_{[1]}^{[m-1,1]}$ is suggested
for all $m$).

\paragraph{Colored HOMFLY for torus knots.}
Especially simple are the HOMFLY polynomials for the torus knots.
In this case,
the coefficients $\K_R^Q$ are known explicitly
in far more generality:
for all torus knots $[m,n]$ \cite{torus}:
\be\label{19}
\K_R^Q = q^{\frac{n}{m}\varkappa_Q}C_R^Q
\ee
where $C_R^Q$ are provided by ``the Adams operation'':
\be\label{20}
S_R(p^{[m]}) = \sum_Q C_R^Q S_Q(p),
\ \ \ \ \ \
p^{[m]}_k = p_{mk}
\ee
and
\be
\varkappa_Q = \nu_{Q'} - \nu_{Q}
\label{evs}
\\
\nu_Q = \sum_i (i-1)Q_i,
\ \ \ \
\varkappa_Q = {1\over 2}\sum_i Q_i(Q_i-2i+1)=\sum_{(i,j)\in Q}(i-j)
\ee
($Q'$ denotes the transposed Young diagram).
Similarly, for $l$-component torus knot the colored HOMFLY polynomials
depend on $l$ different representations, and so do the coefficients
$C_{R_1\ldots R_l}^Q$ so that the Adams operation reads
\be
\prod_{a=1}^l S_{R_a}(p^{[m]}) = \sum_Q C_{R_1\ldots R_l}^Q S_Q(p)
\ee
In fact, for torus knots the $\beta$-deformations of
eqs.(\ref{10}),(\ref{19}),(\ref{20}) are known \cite{DBMMSS} which describe the character
(MacDonald) decomposition of superpolynomials. It is extremely interesting to find
a $\beta$-deformation of (\ref{13})-(\ref{15}).

\section{Integrability}

\paragraph{Continuation from $t^*$ to arbitrary $t$.}
One of the main motivations for representation (\ref{HOexpan})
is a possibility to promote the HOMFLY polynomials to KP $\tau$-functions.
Namely, for
\be
H_R(A) = \sum_Q \K_R^Q S_Q^*
\ee
define
\be
{\cal H}_R\{t\} = \sum_Q \K_R^Q S_Q\{t\}
\ee
with the same coefficients $\K_R^Q$.
Similarly, introduce for a given knot
\be
{\cal H}\{t|\bar t\} = \sum_R {\cal H}_R\{t\} S_R(\bar t) =
\sum_{R,Q} \K_R^Q S_R\{\bar t\}S_Q\{t\}
\ee
and for a given link
\be
{\cal H}\{t|\bar t^{(a)}\} = \sum_{R_1\ldots R_l} {\cal H}_{R_1\ldots R_l}\{t\}
\prod_{a=1}^l S_{R_a}(\bar t^{(a)}) =
\sum_{R,Q} \K_{R_1\ldots R_l}^Q \prod_{a=1}^l S_{R_a}(\bar t^{(a)})S_Q\{t\}
\ee
It turns out that this generating function
is a KP $\tau$-function of $t$-variables,
in the case of torus knots.
Its integrability properties w.r.t. the $\bar t$ variables
remain to be understood.

\paragraph{Torus knots.}
In order to study the torus case, let us note that $\varkappa_Q$ is
the eigenvalue
\be
\hat W_{[2]} S_Q(t) = \varkappa_Q S_Q(t)
\label{caj}
\ee
of the simplest cut-and-join operator $\hat W_{[2]}$ \cite{MMN}
on the Schur eigenfunction $s_Q\{p\}$ corresponding to the Young diagram $Q$.
It is manifestly given by
\be
\hat W_{[2]}={1\over 2}\sum_{a,b}\left[(a+b)p_ap_b{\p\over\p p_{a+b}}+
abp_{a+b}{\p^2\over\p p_a\p p_b}\right]
\ee
Then, using the Cauchy formula
\be
\sum_R S_R\{t\} S_R\{\bar t\} = \exp \sum_k kt_k\bar t_k
\ee
one obtains
\be
{\cal H}^{[m,n]}\{t,\bar t\} =
q^{-\frac{n}{m}\hat W(t)} e^{\sum_k mk t_{mk}\bar t_k}
\ee
The exponential of $t$-variables is the simplest KP $\tau$-function.
Since the cut-and-join operator $\hat W$
is an element of the group $GL(\infty)$,
its action preserves KP-integrability in $t$ \cite{intcharexp,Kaz}.
Therefore for arbitrary torus knot $\l[m,n]$
the generating function ${\cal H}^{[m,n]}\{t,\bar t\}$
is, indeed, the KP $\tau$ function in $t$ (but not in $\bar t$).

Similarly, the generating function of the torus link, ${\cal H}\{t|\bar t^{(a)}\} $
is the same $\tau$-function with redefined parameters $\bar t_k\to
\sum_{a=1}^l\bar t_k^{(a)}$:
\be
{\cal H}^{[m,n]}\{t|\bar t^{(a)}\} =
q^{-\frac{n}{m}\hat W(t)} e^{\sum_k mk t_{mk}\left(\sum_{a=1}^l\bar t_k^{(a)}\right)}
\ee

\paragraph{Non-torus knot/link examples.}

In the case of non-toric knots the same generating function is typically
not a KP $\tau$-function. In order to check this, let us consider the first
non-trivial Pl\"ucker relation (\ref{Plure}) for $g_Q=\sum_Rh^Q_RS_R(\bar t)$
and a 4-strand knot. Then, since
$g_0=1$, $g_{[1]}=g_{[2]}=g_{[11]}=g_{[21]}=0$  in this case, in order to satisfy
(\ref{Plure}), one inevitably should have $g_{[22]}=h^{[22]}_{[1]}=0$.
This is the case for the torus
knots, and not typically the case for others. Indeed, for the first 4-strand knots
from the Rolfsen table (up to 8 crossings)
\cite{katlas} one has \cite{II}:

\vspace{1cm}

\begin{tabular}{|c|c|}
\hline
\hbox{knot}&$h^{[22]}_{[1]}$\\
\hline
$6_1$&$q^{-1}-q^{1}$\\
$7_2$&$-q^{7}+q^{5}-2q^{3}+3q^{1}-3q^{-1}+2q^{-3}-q^{-5}+q^{-7}$\\
$7_4$&$(q-q^{-1})(q^{6}-q^{4}+3q^{2}-1+3q^{-2}-q^{-4}+q^{-6})$\\
$7_6$&$-q^{7}+2q^{5}-3q^{3}+3q^{1}-3q^{-1}+3q^{-3}-2q^{-5}+q^{-7}$\\
$7_7$&$-q^{7}+3q^{5}-4q^{3}+5q^{1}-5q^{-1}+4q^{-3}-3q^{-5}+q^{-7}$\\
$8_4$&$(q-q^{-1})(q^4-q^2+1-q^{-2}+q^{-4})$\\
$8_6$&$(q-q^{-1})(q^2+1+q^{-2})(q^2-1+q^{-2})$\\
$8_{11}$&$-q^{3}+q^{-3}$\\
$8_{13}$&$(q-q^{-1})(q^4-q^2+1-q^{-2}+q^{-4})$\\
$8_{14}$&$(q-q^{-1})(q^2+1+q^{-2})(q^2-1+q^{-2})$\\
$8_{15}$&$(q-q^{-1})(q^{6}-2q^{4}+2q^{2}-3+2q^{-2}-2q^{-4}+q^{-6})$\\
\hline
\end{tabular}

\vspace{1cm}

Thus, for all these knots the Pl\"ucker relation (\ref{Plure}) is not satisfied
(torus knots with 4 strands have more than 8 crossings).

\section{Difference equations for torus knots in the case of $N=2$}

\paragraph{Knot polynomial as an average.}
Difference equations, originally nicknamed non-commutative $A$-polynomials
\cite{Apols}
(they are polynomials in powers of the shift operator,
changing the heights of the rows in Young diagram $R$)
are examples of {\it linear} relations between the HOMFLY
polynomials $H_R$ associated with different Young diagrams $R$.
They play the same role as "the string equations" in matrix model theory
and are presumably a piece of the infinite system of
Virasoro like constraints (recursion relations),
which still remain to be discovered.
They can be used to introduce a spectral curve,
then, the AMM/EO topological recursion \cite{AMMEO} presumably
restores the entire HOMFLY polynomial; by now,
this was checked \cite{DF} for Jones polynomials
in two particular cases of non-torus knots and for the torus knots.

In the previous section, we explained that
the character decompositions provide a natural approach
for the study of quadratic relations.
Now we demonstrate that they are not less useful
for the search of linear relations.
Again, we restrict our consideration to the torus knots,
and also to the case of $SL(2)$ group, i.e. to $A=q^N=q^2$.
In this case, non-vanishing are only the Schur polynomials
associated with the single-row Young diagrams,
\be
S_k[X] = \frac{x_1^{k+1}-x_2^{k+1}}{x_1-x_2} =
x_1^k + x_1^{k-1}x_2 + \ldots + x_2^{k}
\ee
and the two-row diagrams, but the latter ones are expressed
through the previous ones:
\be
S_{[k-l,l]} = S_{k-2l}
\ee
All other
\be
S_{[k-l_1-l_2,l_1,l_2,\ldots]} = 0,
\ \ \ \ \ \ {\rm for}\ l_2\neq 0,\\
\ldots
\ee
Note, however, that
$\varkappa_{[k-l,l]}
= (k-l)(k-l-1) + l(l-3) \neq \varkappa_{k-2l} = (k-2l)(k-2l-1)$.

In fact, below we deal with the characters
of the simple Lie groups ($SU(2)$ in this case),
hence, we slightly rescale the character $S_{[k]}\to \hbox{S}_{[k]}=S_{[k]}/x_2^k$.
This effects just a normalization factor of the knot polynomial.

The difference equation is going to be in the variable $k$, that is,
the height of the single-row Young diagram.
The property which we are going to use in the derivation
of this equation is that the HOMFLY polynomial
$H_R$ for the knot ${\cal K}$ represented as an $m$-strand braid
${\cal B}^{\cal K}$
can be presented as an average over the $N\times N$ matrix
$U=e^u$ of the character $\hbox{S}_R(U^m)$
with some measure which depends on the braid:
\be\label{35}
\hbox{H}_R^{\cal K} = \Big< \hbox{S}_R( U^m) \Big>_{{\cal B}^{\cal K}}
\ee
Since thus presented HOMFLY polynomial has a specific normalization, we denote it
differently.

At least, for the torus knots such a representation does exist,
and is explicitly given, for example, by the
matrix model \cite{BEM}.
We shall use this concrete model to derive an explicit
shape of the difference equation (i.e. of the $A$-polynomial).

The very fact that an equation exists
does not depend on the shape of the measure.
Its {\it raison d'etre} is very simple:
\be
\hbox{H}_{[k+1]}^{\cal K} - \hbox{H}_{[k-1]}^{\cal K} =
\Big< \hbox{S}_{[k+1]}( U^m) - \hbox{S}_{[k-1]}( U^m) \Big>_{{\cal B}^{\cal K}}
= \Big< e^{m(k+1)(u_1-u_2)} + e^{-m(k+1)(u_1-u_2)} \Big>_{{\cal B}^{\cal K}}
= V_k^{{\cal B}_{\cal K}}(q)
\ee
where $V_k$ is a ${\cal B}_{\cal K}$-dependent polynomial in $q$,
which can be explicitly evaluated if the measure is known.

\paragraph{$V_k^{[m,n]}$ from the matrix model.}

According to \cite{BEM},
for the torus knot ${\cal B}_{\cal K} = [m,n]$,
which is represented as an $m$-strand braid,
the measure is given by
\be
\Big< \ldots \Big>^{[m,n]} =\left({\eta\over 2\pi h}\right)^{N/2}
\prod_{i=1}^N \int du_i e^{-\frac{\eta u_i^2}{h}}
\prod_{i<j}\sinh(u_i-u_j)\, \sinh\Big(\eta(u_i-u_j)\Big)
\  \Big(\ldots\Big)
\ee
where $\eta = \frac{m}{n}$ and $q = e^{h}$.

In the case of $N=2$ the average of any exponential of $u_1$ and $u_2$ is
a 4-term polynomial in $q$, in particular\footnote{
The character $S_k$ itself is a sum of $k+1$ terms \cite{Morton}:
$$
\Big< \hbox{S}_k(U^m) \Big>_{[m,n]} =
\Big< \sum_{j=0}^k e^{m(k-2j)(u_1-u_2)}\Big>_{[m,n]} =
$$
$$
=
q^{{m^2+n^2\over 2mn}}\sum_{j=0}^{k-1}  q^{{mn(k-2j)^2}}
\left\{
q\Big(q^{(m+n)(k-2j)} + q^{-(m+n)(k-2j)}\Big)
-{1\over q}q\Big(q^{(n-m)(k-2j)} + q^{(m-n)(k-2j)}\Big)\right\}
$$
},
\be
V_k^{[m,n]} = \Big<  e^{m(k+1)(u_1-u_2)} + e^{-m(k+1)(u_1-u_2)} \Big>_{[m,n]}
= \\ =
q^{{m^2+n^2\over 2mn}}q^{mn(k+1)^2\over 2}
\left\{
q\Big(q^{(m+n)(k+1)} + q^{-(m+n)(k+1)}\Big)
-{1\over q}\Big(q^{(n-m)(k+1)} + q^{(m-n)(k+1)}\Big)\right\}
\label{Vkmn}
\ee
This result coincides with the difference equation for the torus knots obtained in
\cite{Hikami}. However, this second order difference equation reduces to the
first order one for the torus knots of the series $[2,2s+1]$ \cite{Hikami}.
Let us illustrate it in the simplest example of the trefoil.

\paragraph{Simplest example of the trefoil $3_1 = [2,3]$.}

In this case, the Jones polynomial ($l=k+1=2j+1$ for the spin $j$ representation) \cite{Jones}
\be
J_l(q)= [l]_q \left(1+ \sum_{i=1}^{l-1}
(-)^i q^{-i(i+3)}q^{-2il}
\prod_{j=1}^i \Big(1-q^{2(l-j)}\Big)\Big(1-q^{2(l+j)}\Big)\right)
\ee
satisfies the difference equation \cite{Apols} (note that we used in
similar formulas in \cite{3dAGT} different normalization of the Jones polynomials, and also
$q\to 1/q$)
\be
J_l + q^{-3(2l-1)}J_{l-1} =
q^{-3(l-1)}[2l-1]_q
\ee
It follows that
\be
\hbox{H}_k = q^{3(k+1)^2}q^{1/12-2}\{q\}J_{k+1}
\ee
satisfies
\be
\hbox{H}_{k+1} + \hbox{H}_k = q^{3(k^2+3k+3)}[2k+3]_q\{q\}q^{{1\over 12}-2}
\ee
and, taking the difference of two successive equations of this form, one gets
\be
\hbox{H}_{k+1} - \hbox{H}_{k-1} = q^{{1\over 12}+3(k+1)^2+1}\{q\}\left(q^{3(k+1)}[2k+3]_q-
q^{-3(k+1)}[2k+1]_q\right)
\ee
which is exactly $V_k^{[2,3]}$ in (\ref{Vkmn}).

\section{Conclusion}

This paper is devoted to revealing a significance of the character expansion
of the HOMFLY polynomials.
We explained that the character decomposition can be defined
unambiguously for particular braid representations of the knot.
Then, it can be studied in full generality for {\it all} braids
with the particular number $m$ of strands, thus, putting under
control the dependence of the HOMFLY polynomial on the shape of the knot.

$\bullet$ We presented explicit results for $m=2,3,4$ from \cite{II},
demonstrating the existence of an additional universal hierarchical structure
in the formulas.

$\bullet$ It is clear from these examples that the character decomposition
provides explicit formulas for the entire {\it series} of knots
depending on arbitrary parameters,
thus, opening a possibility to study various hidden relations
between knot invariants.

$\bullet$ The character decomposition explicitly separates dependencies
on the shape of the knot and the size $N$ of the group.
This allows one to continue the formulas from the particular "frozen"
values of the hidden time-variables $p_k=p^*_k = \frac{A^k-A^{k}}{q^k-q^{-k}}$
to arbitrary values of $p_k$, a trick which already proved extremely
useful in the study of the torus {\it super}polynomials \cite{DBMMSS}.

An open question remains about the further separation of the
$R$-variable (labeling the representation).

\bigskip

These results are immediately applicable to the search
of linear and quadratic relations, in particular, of integrability
properties of the HOMFLY polynomials and of difference equations
("A-polynomials"), which they satisfy.
We explicitly demonstrated such applications in the case
of the torus knots and showed that:

$\bullet$ When continued to arbitrary values of $p_k$,
these polynomials become KP $\tau$-functions;
this is a non-trivial property, literally correct only
for the torus knots.

$\bullet$ Difference equations for the colored Jones polynomials
can be derived "in one line" for the arbitrary
torus knot $[m,n]$.
Moreover, generalizations to HOMFLY in this case are
also straightforward.

\bigskip

It would be interesting to find
an extension of the matrix model including the time-variables
$\{p_k\}$. It should be straightforward, since the $W$-representation is
known and, thus, the methods of \cite{MShWrep} can be applied.
One can also pose the question about a complete system of Virasoro like
constraints, of which the difference equation should be just the
single (lowest?) constituent.

Also of interest is search for a counterpart of (\ref{35}) for non-torus knots or
there can be obstacles for existence of measures with such a property.

\bigskip

The most intriguing is an unambiguous definition of the
character decomposition after the $\beta$-deformation
from HOMFLY to the {\it super}polynomials.
In the case of the torus knots it is recently found in
\cite{DBMMSS},
generalization to arbitrary knots
is the next point on agenda.

\section*{Acknowledgements}

Our work is partly supported by Ministry of Education and Science of
the Russian Federation under contract 14.740.11.0608, by RFBR
grants 10-02-00509 (A.Mir.), 10-02-00499 (A.Mor.) and 11-02-01220 (And.Mor.),
by joint grants 11-02-90453-Ukr, 09-02-93105-CNRSL, 09-02-91005-ANF,
10-02-92109-Yaf-a, 11-01-92612-Royal Society.

\end{document}